\documentclass[12pt]{iopart}

\usepackage{iopams}

\expandafter\let\csname equation*\endcsname\relax

\expandafter\let\csname endequation*\endcsname\relax

\usepackage{amsmath}

\usepackage{graphicx}

\usepackage[usenames]{color}

\newcommand{\be}{\begin{equation}}
\newcommand{\ee}{\end{equation}}
\newcommand{\bea}{\begin{eqnarray}}
\newcommand{\eea}{\end{eqnarray}}

\definecolor{bc}{rgb}{0, 0.7, 0.0}

\newcommand{\RefA}[1]{{\color{black}#1}}
\newcommand{\RefB}[1]{{\color{black}#1}}
\newcommand{\RefC}[1]{{\color{black}#1}}

\newcommand{\ud}{\mathrm{d}}
\newcommand{\LCm}{{\scriptscriptstyle -}} %LC supersripts
\newcommand{\LCp}{{\scriptscriptstyle +}}

\newcommand{\LCperp}{{\scriptscriptstyle \perp}}

\makeatletter
\def\footnoterule{\kern-3\p@
  \hrule \@width \textwidth \kern 2.6\p@} % the \hrule is .4pt high
\makeatother

\begin{document}

\title[Superintegrable relativistic systems]{Superintegrable relativistic systems in scalar background fields}

\author{L.~Ansell, T.~Heinzl and A.~Ilderton}
\address{Centre for Mathematical Sciences, University of Plymouth, PL4 8AA, UK}
\eads{\mailto{lauren.ansell@plymouth.ac.uk}, \mailto{thomas.heinzl@plymouth.ac.uk}, \mailto{anton.ilderton@plymouth.ac.uk}}
\vspace{10pt}

\begin{abstract}
We consider a relativistic charged particle in a background scalar field depending on both space and time. Poincar\'e, dilation and special conformal symmetries of the field generate conserved quantities in the charge motion, and we exploit this to generate examples of superintegrable relativistic systems. We also show that the corresponding single-particle wavefunctions needed for the quantum scattering problem can be found exactly, by solving the Klein-Gordon equation. 
\end{abstract}

\vspace{2pc}

\clearpage
\section{Introduction}
A classical dynamical system with $2n$-dimensional phase space is superintegrable if it admits $n+k$ functionally independent conserved quantities $Q_j$ on phase space, with $1 \leq k \leq n-1$, of which $n$ are in involution, i.e.~their Poisson brackets obey $\{Q_i,Q_j\} = 0\, \forall \, i,j = 1,...,n$~\cite{Wojciechowski:1983}. For autonomous systems the Hamiltonian itself may be taken as one of the $Q_j$. The system is called maximally superintegrable if $k=n-1$ and  minimally superintegrable if  $k = 1$.

While most known superintegrable systems correspond to non-relativistic physics, i.e.~describe dynamics on \RefC{two- or three-dimensional Euclidean space}~\cite{Miller:2013}, our interest here is in identifying relativistic superintegrable systems, continuing the programme started in~\cite{Heinzl:2017blq}. Previously we considered classical particles interacting with electromagnetic backgrounds, and showed that if the \textit{background} possesses a Poincar\'e symmetry then there is automatically a conserved quantity in the \textit{particle} motion. Using this we found examples of superintegrable systems in which all conserved quantities corresponded to Poincar\'e symmetries of the background, and examples in which some quantities corresponded to non-Poincar\'e symmetries on phase space. 

Here we will consider a relativistic particle interacting with a scalar, rather than electromagnetic, field (which represents an early model of gravity~\cite{Nord,EF}). This simplified setting has the advantage that it allows us to go beyond Poincar\'e symmetries and exploit dilation and special conformal transformations in the construction of superintegrable systems, unlike for electromagnetic, or vector, backgrounds.

This paper is organised as follows. In Sect.~\ref{SECT:REVIEW} we describe the relativistic dynamics of a particle in a scalar background, equivalently a particle with a position-and-time-dependent mass. We show that conformal symmetries of this mass imply conserved quantities in the particle motion. Based on this, we present in Sect.~\ref{SECT:EG} a series of minimally and maximally superintegrable relativistic systems. In Sect.~\ref{SECT:KG} we take a step toward the quantum problem, \RefB{investigating how, given a spacetime-dependent mass, the classical conserved quantities enter in the solution of the corresponding Klein-Gordon equation for a scalar field.} We conclude in Sect.~\ref{SECT:CONCS}

\section{Dynamics of a point charge in a scalar field}\label{SECT:REVIEW}
%%%%%%%
%
The action of a relativistic particle of rest mass $m_0$ in a scalar background field~$V(x)$ is, see e.g.~\cite{Straumann:2000ke},
\be \label{ACTION}
	S = -\int\!\ud \tau \; \big( m_0 + V(x)\big) \RefC{\sqrt{\eta_{\mu\nu}{\dot x}^\mu {\dot x}^\nu}}\;,
\ee
in which $x^\mu \equiv x^\mu(\tau)$, with $\tau$ (the proper time) parameterising the worldline, $\dot{x}^\mu\equiv \ud x^\mu /\ud \tau$ and \RefC {$\eta_{\mu\nu} = \text{diag}(1,-1,-1,-1)$ the Minkowski metric.}   \RefB{(The coupling between the particle and scalar field, traditionally denoted~$e$, is absorbed into $V$ throughout.)}  The scalar field couples to the particle like a spacetime-dependent mass, so we write
\be
	m_0 +V(x) \equiv m(x) \;,
\ee  
from here on. We  refer to $m(x)$ as a dynamical mass. \RefC{The action (\ref{ACTION}) is equivalent to that of a test particle in a curved spacetime with a conformally flat metric
\be\label{G}
	g_{\mu\nu}(x) = \frac{m^2(x)}{m_0^2} \eta_{\mu\nu} \;,
\ee
where the dynamical mass appears as a conformal factor. This represents an old-fashioned ``scalar'' model of gravity~\cite{Nord,EF,Straumann:2000ke}, see also~\cite{vanHolten:2006xq}.} Varying the action (\ref{ACTION}) yields the Euler-Lagrange equations, 
\RefA{
\be \label{LORENTZ-0}
	\frac{\ud}{\ud\tau}\bigg(\frac{m \dot{x}_\mu}{\sqrt{\dot{x}.\dot{x}}}\bigg) = \sqrt{\dot{x}.\dot{x}}\, \partial_\mu m \;.
\ee
Expanding out and contracting with $\dot{x}^\mu$ yields
\be\begin{split}
	\frac{\dot{x}^\mu}{\sqrt{\dot{x}.\dot{x}}}  \frac{\ud}{\ud\tau}\bigg(\frac{\dot{x}_\mu}{\sqrt{\dot{x}.\dot{x}}}\bigg) = 0 \;, 
\end{split}
\ee
which, integrating up, implies that $\dot{x}^2 = $ constant, hence the particle is on-shell. We fix $\dot{x}^2=1$ from here on.}  The equations of motion reduce to 
\be \label{EOM-ENKEL}
	\frac{\ud}{\ud\tau}\big(m \dot{x}_\mu\big) = \partial_\mu m \;,
\ee	
which may be regarded as a force law $m \ddot{x}_\mu = \big(\RefC{\eta_{\mu\nu} - \dot{x}_\mu \dot{x}_\nu} \big) \partial^\nu m$, with the right-hand side replacing the Lorentz force of the vector case, and the tensor structure guaranteeing orthogonality of velocity and acceleration, $\dot{x}. \ddot{x} = 0$, hence the constancy of $\dot{x}^2$. \RefC{The equations of motion are equivalent to the geodesic equations in the metric (\ref{G}). For approaches to integrability based on geodesic flows see~\cite{Calogero}.}

\subsection{The conformal group and conserved quantities}
We wish to identify when a symmetry of the \textit{background}, or dynamical mass, automatically implies the existence of a conserved quantity in the \textit{particle} motion. To do so we need the canonical momenta $p_\mu$ following from (\ref{ACTION}), which are
\be\label{PMX}
	p_\mu = m(x) \dot{x}_\mu \;,
\ee
and which obey the ``dynamical mass-shell constraint'' $p.p = m^2(x)$. Now let $\xi^\mu(x)$ be a vector field defining the infinitesimal form of some coordinate transformation, and define $Q := \xi.p$. Then one can show directly from the equations of motion (\ref{EOM-ENKEL}) that
\be\label{invarians}
	2 m(x) \frac{\ud Q}{\ud \tau} = \mathcal{L}_\xi m^2 +  p^\mu p^\nu \big( \partial_\mu \xi_\nu + \partial_\nu \xi_\mu\big) \;,
\ee
in which $\mathcal{L}_\xi = \xi.\partial $ is the Lie derivative of any scalar quantity. For $Q$ to be conserved we need the \RefC{right hand side} of (\ref{invarians}) to vanish, and we demand that it does so through properties of the field and the transformation, not the details of the orbit. It is worth pointing out that the analogous equation in the electromagnetic (vector) case has essentially the same \RefC{right hand side} as (\ref{invarians}), except that the Lie-derivative term has an extra power of $p$, hence the two terms must vanish individually. The situation here is different; the most general way to kill the \RefC{right hand side} of (\ref{invarians}) includes contracting $p^\mu p^\nu$ with the metric tensor, so that it can be replaced by $m^2(x)$, and then only the \textit{sum} of the two terms need vanish. This means $\xi$ must obey
\be\label{CKE}
	\partial_\mu \xi_\nu + \partial_\nu \xi_\mu \propto \eta_{\mu\nu} \implies \qquad \partial_\mu \xi_\nu + \partial_\nu \xi_\mu = \frac{1}{2} \eta_{\mu\nu}\partial\cdot \xi \;,
\ee
in which the scalar factor on the \RefC{right hand side} was determined by taking the trace. This is nothing but the conformal Killing equation, with the 15-parameter solution
\be\label{killing-sol}
	\xi_\mu(x) = a_\mu + \omega_{\mu\nu} x^\nu + \lambda x_\mu + c_\mu x^2-2(c.x) x_\mu \;, \qquad (\omega_{\mu\nu} = - \omega_{\nu\mu}) \;,
\ee
describing, respectively, translations, Lorentz transformations, dilations and special conformal transformations, spanning the conformal group. For these transformations~(\ref{invarians}) becomes
\be\label{invarians2}
	2 m(x) \frac{\ud Q}{\ud \tau} = \mathcal{L}_\xi m^2 +  m^2 \frac{1}{2}\partial\cdot\xi \;,
\ee
and it follows that there is a conserved quantity $\xi.p$ in particle motion when the dynamical mass obeys
\be\label{L-xi-general}
	\mathcal{L}_\xi m^2 + m^2 \frac{1}{2}\partial\cdot\xi = 0 \;.
\ee
For Poincar\'e transformations (with $\partial.\xi=0$) this says that the dynamical mass must be symmetric under the transformation, $\mathcal{L}_\xi m^2=0$, while for dilations and special conformal transformations the mass must transform with a weight. (\RefC{The appearance of the conformal group is completely natural, as it is the isometry group of the metric~(\ref{G}).})

We will exploit these results below to \textit{construct} systems which have sufficiently many conserved phase space quantities $Q\equiv Q(x^\mu,p_\mu)$ to be superintegrable. To formalise this, though, we need a Hamiltonian\footnote{We comment that the Poisson bracket any two quantities $Q_j=\xi_j(x).p$ is 
\be
	\{Q_1,Q_2\} = \big( \xi_2.\partial \xi_1^\nu - \xi_1.\partial \xi_2^\nu \big)p_\nu  \;,
\ee
and the commutator of the two transformations acting on any scalar field is
\be
	[\mathcal{L}_{\xi_2}, 	\mathcal{L}_{\xi_1}] = \big( \xi_2.\partial \xi_1^\nu - \xi_1.\partial \xi_2^\nu \big) \partial_\nu \equiv \mathcal{L}_{[\xi_2,\xi_1]} \;,
\ee
and hence we see that two conserved quantities are in involution provided that the associated Poincar\'e generators commute.}; we therefore now briefly review the Hamiltonian approach to relativistic mechanics.

%%%%%%%%%%%%%%%%%%%
\subsection{Hamiltonian formulation}
%%%%%%%%%%%%%%%%%%%
%
Euler's homogeneous function theorem, here in the guise of reparametrisation invariance, means that the Hamiltonian corresponding to (\ref{ACTION}) vanishes. \RefC{There are several ways to tackle this apparent problem. It is possible to retain manifest covariance by working with a second-order energy functional, $L \sim g_{\mu\nu} \dot{x}^\mu \dot{x^\nu}$ \cite{Wald:1984,Hughston:1990,Straumann:2013}, rather than the homogeneous length functional or action (\ref{ACTION}). Minimising the former reproduces the equation of motion (\ref{EOM-ENKEL}), hence geodesic motion in the metric (\ref{G}). Alternatively, in order to e.g.~make more explicit contact with the bulk of the superintegrability literature which focusses on non-relativistic systems, one can sacrifice manifest covariance by following Dirac~\cite{Dirac:1949cp}: rather than using $\tau$ as the time parameter, one can choose a physical time, which is a function of the $x^\mu$. Relativistic covariance implies that there is no unique choice of time, and while all are ultimately equivalent one choice or another may have advantages in particular situations. (Indeed, the introduction of a nontrivial background field automatically breaks \textit{manifest} Lorentz invariance, and background field problems can often be simplified by an appropriate choice of co-ordinates, or time variable.)} Each choice of time comes with its own set of (six) phase space variables and a Hamiltonian given by a particular component of $p_\mu$, found by rearranging the dynamical mass-shell constraint $p.p=m^2(x)$.  Below we describe the two choices needed for this paper. For reviews and references see~\cite{Heinzl:2017blq, Heinzl:2000ht}.
In the ``instant form'', time is $t$ while six-dimensional phase space is spanned by the co-ordinates $\mathbf{x}=(x^j)=(x,y,z)$ and their conjugate momenta, $\mathbf{p}=(p_j)=(p_1,p_2,p_3)$.  The Hamiltonian is  
\be\label{H-INST}
	H \equiv p_0 = \sqrt{{\bf p}^2 + m^2(t,{\bf x})} \;,
\ee
and may be explicitly time-dependent, through $m^2$. The time evolution of any quantity $Q$ is determined by
\be
	\frac{\ud Q}{\ud t} = \frac{\partial Q}{\partial t} - \{Q,H\} \;,
\ee
where the Poisson bracket is
\be
	\{X,Y\} := \frac{\partial X}{\partial x^j}\frac{\partial Y}{\partial p_j} -\frac{\partial X}{\partial p_j}\frac{\partial Y}{\partial x^j}  \;.
\ee
\RefB{The instant form is convenient for discussing the non-relativistic limit, in which the particle velocity obeys $\ud x^j/\ud t\ll 1$. This limit is taken, recalling (\ref{PMX}), by extracting the factor of $m^2$ from inside the square root of (\ref{H-INST}) and expanding  in powers of ${\bf p}^2/m^2$. The result is
\be\label{HNR}
	H \to H_\text{non rel.} = \frac{{\bf p}^2}{2m(t,{\bf x})} + m(t,{\bf x}) \;.
\ee
This is exactly the form considered in investigation of \textit{non-relativistic} superintegrable systems with dynamical mass, see e.g.~\cite{Ball,Nik,Nik2,Ghose}. This gives an explanation for why the symmetries of the non-relativistic system (\ref{HNR}) are given in terms of 3-d conformal Killing vectors~\cite{Nik}; the non-relativistic system is the limit of our relativistic system, the dynamics of which is equivalent to that in a nontrivial metric~(\ref{G}), the isometry group of which is the 4-d conformal Killing group.}

\RefB{In what follows, we will also use the ``front form'' of dynamics, in which} time is $x^\LCp\equiv t+z$, and phase space is spanned by the `longitudinal' coordinate $x^\LCm\equiv t-z$, the `transverse' coordinates ${\boldsymbol x}^\LCperp = (x^\LCperp) = (x,y)$, and their conjugate momenta $p_\LCm$ and ${\bf p} = (p_\LCperp) = (p_1,p_2)$ respectively. The Hamiltonian and Poisson bracket are  (summation convention is used throughout for the index $\perp$)
\be\label{H-LF}
	H \equiv p_\LCp = \frac{{\bf p}_\LCperp {\bf p}_\LCperp +m^2(x^\mu)}{4 p_\LCm} \;,
\ee
\be\label{PB-LF}
	\{A,B\} = 
	\frac{\partial A}{\partial x^\LCm}\frac{\partial B}{\partial p_\LCm} 
	 -
	 \frac{\partial A}{\partial p_\LCm} \frac{\partial B}{\partial x^\LCm}
	+
	 \frac{\partial A}{\partial x^\LCperp}\frac{\partial B}{\partial p_\LCperp} 
	 -
	  \frac{\partial A}{\partial p_\LCperp} \frac{\partial B}{\partial x^\LCperp}  \;,
\ee
and the equation of motion for any quantity $Q$ is now
\be\label{utveckling-lf}
	\frac{\ud Q}{\ud x^\LCp} = \frac{\partial Q}{\partial x^\LCp} - \{Q,H\} \;.
\ee

%%%%%%%%%%%%%%%%%%%
\section{Examples}\label{SECT:EG}
%%%%%%%%%%%%%%%%%%%
%
In this section we present a series of superintegrable systems constructed by exploiting the symmetries of the conformal group (\ref{killing-sol}). We begin with illustrative examples in which the dynamical mass is a function of a single spacetime variable. Relativistic covariance then tells us that there can be only three distinct cases, when the chosen spacetime direction is spacelike, e.g.~$z$, timelike e.g.~$t$, or lightlike, e.g.~$x^\LCp$. Following this we give an example superintegrable system with special conformal symmetry. Our examples will also illustrate symmetries on phase space, the role of boundary conditions, and equivalent autonomous systems.

\subsection{The spacelike case}

Consider a dynamical mass $m^2=m_0^2+B(z)$. This is the scalar analogue of a position-dependent magnetic field~\cite{Marchesiello:2015,Heinzl:2017blq}. Using the instant form, the Hamiltonian is
\be\label{HB}
	H = \sqrt{{\bf p}^2 + m_0^2 + B(z)} \;.
\ee
Clearly $p_1$, $p_2$ and $H$ are (independent) conserved quantities. To search for others, we follow~\cite{Fris:1965,Marquette:2008,Miller:2013} and make the ansatz that the remaining conserved quantities are polynomials in $p_1$ and $p_2$. The simplest case is to make a linear ansatz, writing
\be\label{ansatz}
	Q = f_1({\bf x},p_3)p_1 +  f_2({\bf x},p_3)p_2 + f_3({\bf x},p_3) \;.
\ee
We then calculate $\ud Q/\ud t$, write out the resulting expression in powers of $p_1$ and $p_2$ and demand that each term vanishes. This yields a series of algebraic or differential equations which determine the functions $f_1$, $f_2$ and $f_3$ and so on. It may be that no conserved quantities are found, in which case one can try again with an ansatz quadartic in momenta, or cubic, and so on~\cite{Marquette:2008,Miller:2013,March2018}.

\begin{figure}[t]
\centering\includegraphics[width=0.6\textwidth]{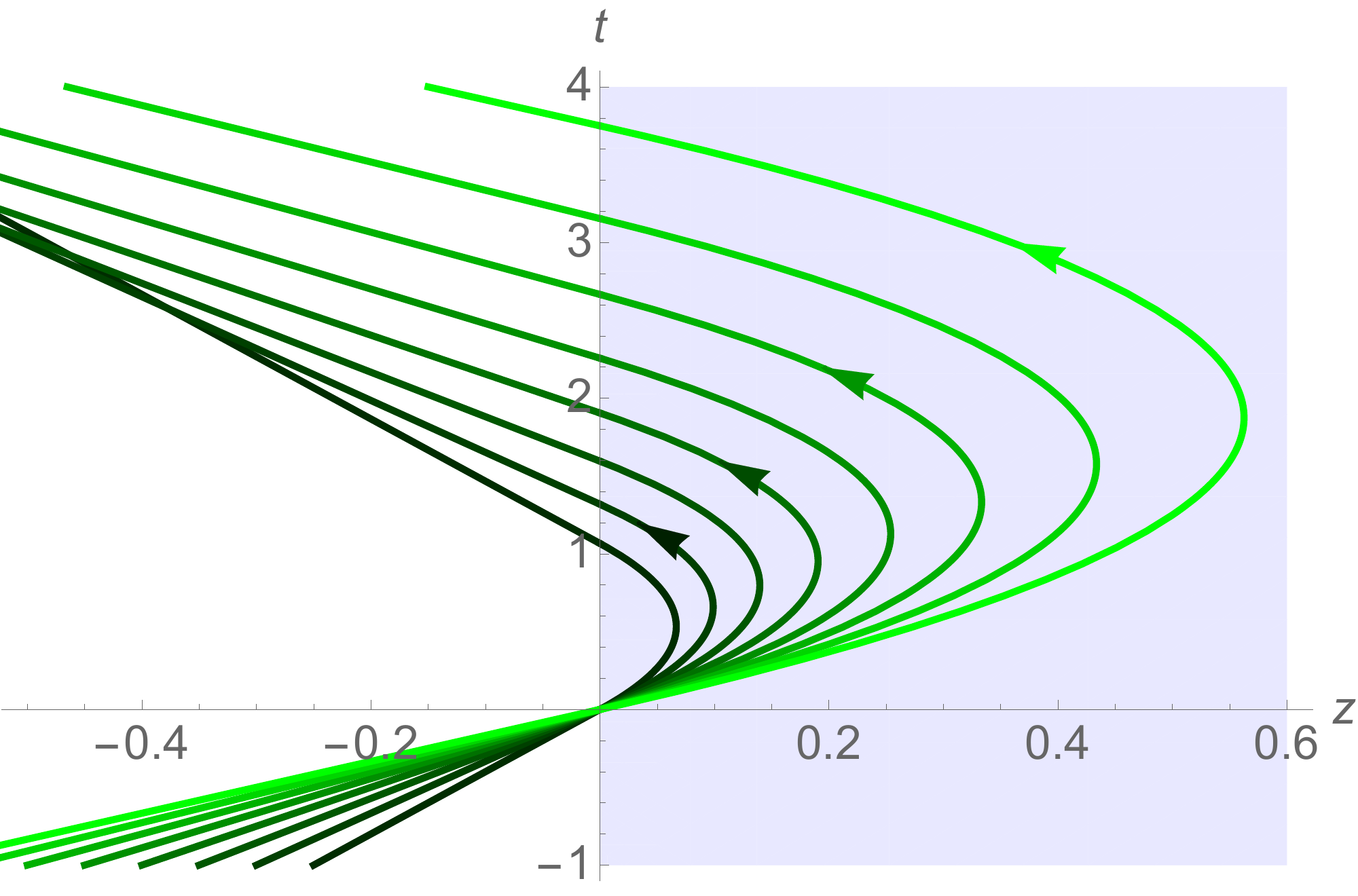}
\caption{\label{FIG:INSTANT} Orbits in the autonomous system with dynamical mass $m^2=m_0^2 +B z$. The particles enter the field (shaded) at at ${\bf x}=0$ at time $t=0$, and with $p_1=p_2=0$. Orbits are shown for $B/m_0^2=1$ and initial momenta $p_3(0)/m_0 \in \{1/4 \ldots 3/5\}$. The general behaviour is that a particle penetrates a certain distance into the region $z>0$ before being turned around and pushed out of the field.}\end{figure}

We illustrate with the simplest nontrivial example, choosing $m^2(t,{\bf x}) =m_0^2 + B z$.  In this case we find that the linear ansatz (\ref{ansatz}) turns out to be sufficient; we find that the $f_j$ can be expressed in terms of a four-parameter family of elementary functions, yielding four conserved quantities (two of which are $p_1$ and $p_2$). Together with the Hamiltonian, this gives us five conserved quantities
\be\label{bevarade-magnetiskt}
	Q_1 = p_1 \;, \quad Q_2 = p_2 \;, \quad Q_3 = 2p_1 p_3 + B x \;, \quad Q_4 = 2p_2 p_3 + B y \;, \quad Q_5 = H \;,
\ee
where $\{Q_1,Q_2,Q_5\}$ are in involution, giving integrability.  \RefB{Note that $\{Q_3,Q_4\}$ do not correspond to elements of the conformal group, see also below, but rather represent `hidden symmetries' on phase space.} Defining $\mathcal{F}=(Q_1,\ldots Q_5)$ and following~\cite{Miller:2013}, the five quantities (\ref{bevarade-magnetiskt}) are functionally independent if the $5\times 6$ matrix   
\be
	\mathcal{M} := \bigg( \frac{\partial \mathcal{F}_l}{\partial x^a} , \frac{\partial \mathcal{F}_l}{\partial p_a} \bigg) \;,
\ee
for $a\in \{1,2,3\}$, has rank 5. For the purposes of presentation we calculate $\mathcal{M}$ using the equivalent set of conserved quantities $\mathcal{F}=(Q_3/B,Q_4/B,Q_5^2/B,Q_1,Q_2,)$, for then
\be
\mathcal{M} = \left(
\begin{array}{cccccc}
 1 & 0 & 0 & \frac{2 p_3}{B} & 0 & \frac{2 p_1}{B} \\
 0 & 1 & 0 & 0 & \frac{2 p_3}{B} & \frac{2 p_2}{B} \\
 0 & 0 & 1 & \frac{2 p_1}{B} & \frac{2 p_2}{B} & \frac{2 p_3}{B} \\
 0 & 0 & 0 & 1 & 0 & 0 \\
 0 & 0 & 0 & 0 & 1 & 0 \\
\end{array}
\right) \;,
\ee
which is upper triangular with rank 5. Hence the system is maximally superintegrable. The solution of the equations of motion proceeds as follows. $p_1$ and $p_2$ are constant, and the Hamiltonian equation of motion for $p_3$ is trivial because $H$ is conserved;
\be
	\frac{\ud p_3}{\ud t} = \frac{B}{2H} = \frac{B}{2Q_5} \implies p_3(t) = p_3(0) + \frac{Bt}{2 Q_5} \;.
\ee
From here the coordinates ${\bf x}(t)$ follow, algebraically, from rearranging (\ref{bevarade-magnetiskt}):
\be
	x(t) = \frac{Q_3 - 2 Q_1 p_3(t)}{B} \;, \quad y(t) = \frac{Q_4 - 2 Q_2 p_3(t)}{B} \;, \quad z(t) = \frac{Q_5^2-Q_\LCperp^2-m_0^2-p_3^2(t)}{B} \;.
\ee
In order to have physical `scattering' boundary conditions, we consider the case where the field `switches on' (and ideally off, which is a simple extension of what follows). We redefine the dynamical mass to obey
\be
	m^2(t,{\bf x}) = \begin{cases} m_0^2 & z < 0 \\
	m_0^2+B z & z\geq 0 \end{cases} \;,
\ee
and consider the motion of particles which reach the interface $z=0$ (from $z<0$) at, without loss of generality, $t=0$. Motion for $t<0$ is free, so that we may specify the initial momentum at $t\leq 0$. The initial data at $t=0$ then fixes the values of $Q_1\ldots Q_5$ and $p_3(0)$, above. Examples of the orbits are plotted in Fig.~\ref{FIG:INSTANT}.

\RefB{Before moving on, we consider again the non-relativistic limit, in which the example above should reduce to case 10 in table 2 of~\cite{Nik}. In that table \textit{three} integrals of motion are identified, belonging to the non-relativistic limit of the conformal group (\ref{killing-sol}), and corresponding in our notation to $p_1$, $p_2$ and $L_z = xp_2 - y p_1$.  These are recovered by our $Q_1$, $Q_2$ and the combination
\be
	\tilde{Q}_3 := Q_3 Q_2 - Q_4 Q_1 = B L_z \;.
\ee
The Hamiltonian is conserved in both the relativistic and non-relativistic theories, so in both cases we may write $\{Q_1,Q_2,H,\tilde{Q_3}\}$ as our four independent quantities (for the appropriate $H$), the first three being in involution. In the relativistic case the fifth quantity can be either of $\{Q_3,Q_4\}$, as above, or e.g.~$Q_3 Q_2 + Q_4 Q_1$ which is \textit{cubic} in momenta~\cite{Marquette:2008}. None of these are conserved in the non-relativistic limit. 
}
\subsection{The timelike case}
%
%%%%%%%%%%%%
We now take $m^2=m_0^2+E(t)$ for $t\geq 0$. The analogous electromagnetic case is a time-dependent electric field. The Hamiltonian is now explicitly time-dependent,
\be\label{HE}
	H(t) = \sqrt{{\bf p}^2 + m_0^2 + E(t)} \;,
\ee
and no longer conserved. On the other hand, since the background is position independent, all three momenta are conserved (and in involution). Because the background is scalar, all three components of angular momentum are also conserved. Taking two of these (the angular momenta $L_j$ obey $p_j L_j=0$, hence not all three are independent) along with the momenta gives five independent conserved quantities. The equations of motion are trivially solved;
\be
	\frac{\ud x^j}{\ud t} = -\frac{p_j}{H(t)} \implies {\bf x}(t) = {\bf x}(0) - {\bf p}\int\limits_0^t\!\ud s \frac{1}{\sqrt{{\bf p}^2+m_0^2 + E(s)}} \;.
\ee
Strictly, only one integral needs to be performed, to find e.g.~the first component of ${\bf x}$, for then the conservation of the angular momenta allows one to write down the remaining co-ordinates algebraically.
%
%%%%%%%
\subsection{The lightlike case}
Now we take $m^2\equiv m^2(n.x)$, where $n^2=0$. This is a scalar plane wave. \RefC{The exact solvability of motion in plane wave backgrounds is well known. The background may be scalar (above) vector (electromagnetism) or tensor (gravity), and all these cases represent superintegrable systems. For the electromagnetic case see~\cite{Heinzl:2017zsr} and for gravity see~\cite{Ilderton:2018lsf} and references therein.}  We use the front form, so
\be \label{H.PW}
  H   = \frac{p_\LCperp^2 + m^2(n.x)}{4p_\LCm} \; .
\ee
There are two ways to proceed. If we choose $n.x=x^\LCm$, then the Hamiltonian is time independent. Clearly $p_\LCperp$ and $H$ are conserved and in involution, corresponding to translation invariance in three dimensions. Plane waves are also invariant under null rotations (see the appendix)~\cite{Heinzl:2017blq,Adamo:2017nia}, giving the corresponding conserved quantities
\be\label{Q.PW.M}
	Q_\LCperp = 2 H x^\LCperp + x^\LCm p_\LCperp \;.
\ee
There are thus five conserved quantities all following from the Poincar\'e symmetries of a plane wave, and the system is maximally superintegrable.

In field theory applications, it is often more convenient to take the dependence of the plane wave to coincide with the choice of time, so let $m^2\equiv m^2(x^\LCp)$. In this case the system is non-autonomous, as $H$ depends on $x^\LCp$, but now all three momenta are conserved, and in involution. In order to work with an (alternative) autonomous system, we can enlarge phase space to eight dimensions with $x^\LCp$ as an additional coordinate, conjugate momentum $p_\LCp$, and a new Hamiltonian $K= H - p_\LCp$~\cite{Arnold}. The time-derivative of any quantity $Q$ is
\be
	\dot{Q} = -\{Q,K\}_* \quad \text{where} \quad \{A,B\}_* = \frac{\partial A}{\partial x^\mu}\frac{\partial B}{\partial p_\mu} - \frac{\partial B}{\partial x^\mu}\frac{\partial A}{\partial p_\mu} \text{ for } \mu \in \{+,-,\perp\}\;.
\ee 
Note that the new time does not appear explicitly, and $\dot{x}^\LCp = -\partial K/\partial p_\LCp = 1$. From here one can verify that the five conserved quantities following from the invariance of the plane wave under translations and null rotations are
\be
	Q_1 = p_1\;, \quad Q_2 = p_2 \;, \quad Q_3 = p_\LCm \;, \quad Q_4 = 2 x p_\LCm + x^\LCp p_1 \;, \quad Q_5 = 2 y p_\LCm + x^\LCp p_2 \;.
\ee
There are a further two conserved quantities involving the extended phase space variables; one is by construction the new Hamiltonian, or equivalently
\be\label{Q6}
	Q_6 = 4p_\LCp p_\LCm - p_\LCperp p_\LCperp - m^2(x^\LCp) \;,
\ee
which is quadratic in the momenta and encodes the dynamical mass-shell constraint\footnote{\RefC{Following a manifestly covariant approach from the beginning, with an affine worldline parameter playing the role of time, would have had the effect of automatically extending phase space to eight dimensions for all our examples, with the dynamical mass-shell condition, e.g.~(\ref{Q6}), appearing as a constraint.}}. The final conserved quantity is
\be
	Q_7 = 4 p_\LCm^2 x^\LCm - p_\LCperp p_\LCperp x^\LCp - \int\!\ud x^\LCp \, m^2(x^\LCp) \;,
\ee
which, in the original phase space, immediately gives the solution to the equations of motion for $x^\LCm$. We have seven globally defined (and independent) conserved quantities which are polynomial in the momenta, and the set $\{Q_1,Q_2,Q_3,Q_6\}$ is in involution. Thus we have a polynomially maximally superintegrable system~\cite{Miller:2013}.

\subsection{Special conformal transformations}
We consider the special conformal transformation $\xi^\mu_c$ generated by $c^\LCm = 1$ (and all other components vanishing). Any function of the form
\be\label{KONF-MASSA}
	m^2(x^\mu) = \frac{1}{x^{\LCp2}} f\bigg(x^\LCm - \frac{x^\LCperp x^\LCperp}{x^\LCp} \bigg) \;,
\ee
obeys the relation (\ref{L-xi-general}) for the transformation $\xi_c$, and is symmetric under three Poincar\'e transformations, namely rotations in the $x$--$y$ plane and two null rotations. Going to the enlarged phase space, we can identify the following five conserved quantities:
\be
	Q_\LCperp = 2 p_\LCm x^\LCperp + x^\LCp p_\LCperp\;, \quad Q_3 = \xi_c.p \;,  \quad Q_4 = x p_2 - y p_1 \;, \quad Q_5 = K\;,
\ee
These are independent and the set $\{Q_1,Q_2,Q_3,Q_5\}$ are four quantities in involution. Hence we have (at least) \textit{minimal} superintegrability in this case. 

\begin{figure}[t!]
\begin{minipage}[t]{0.5\textwidth}
\vspace{0pt}
	\centering\includegraphics[width=\columnwidth]{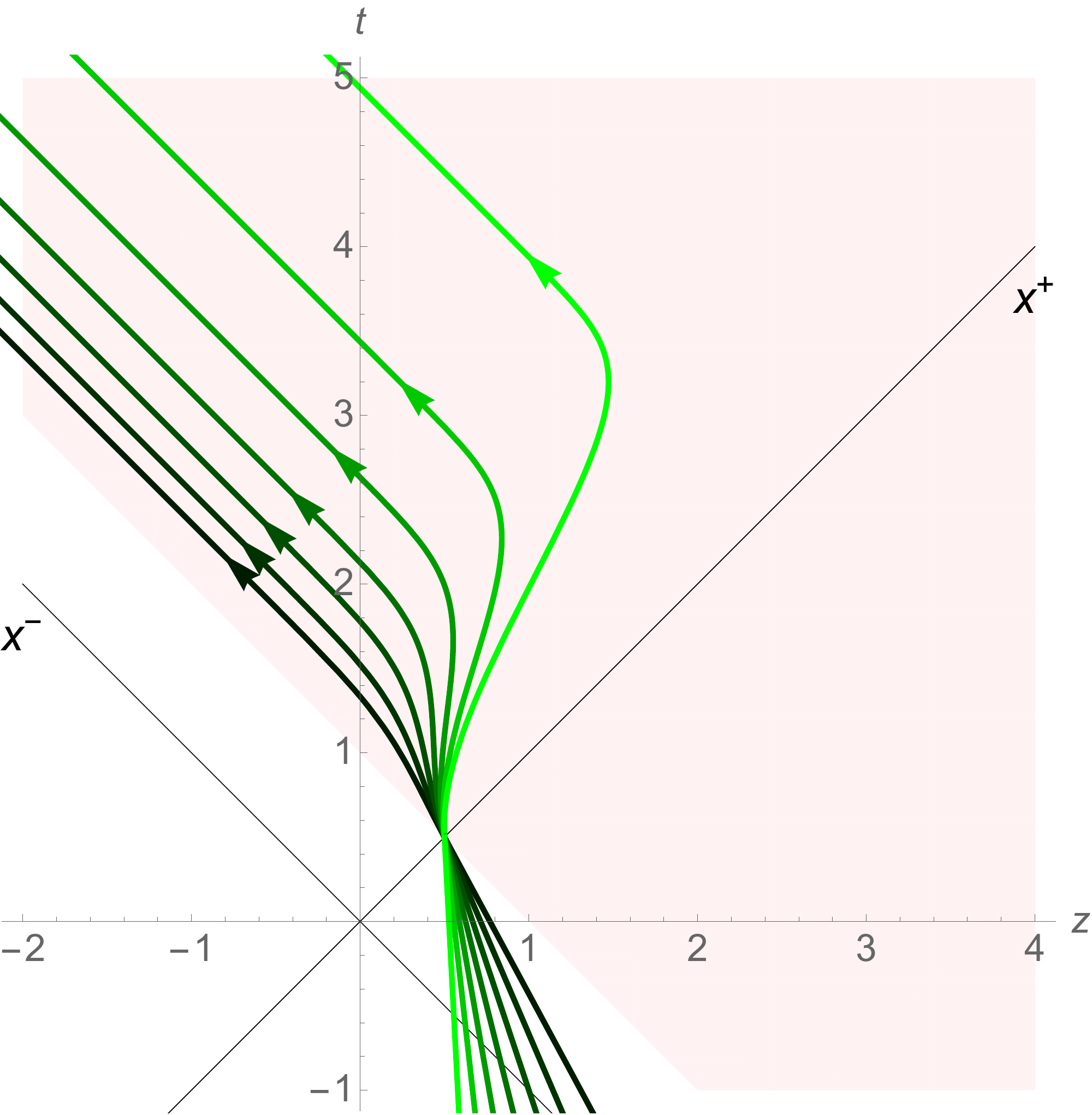}
\end{minipage}
	\hspace{-70pt}
\begin{minipage}[t]{0.65\textwidth}
\vspace{0pt}
\caption{\label{FIG:KONF} For the dynamical mass (\ref{m-def-konf}) the equations of motion admit the partial solution $x^\LCperp = p^\LCperp =0$, which we consider for simplicity. Measuring $x^\LCp$ and $x^\LCm$ in units of $L$ and $1/k$ respectively, the nontrivial part of the orbit is
	\be
		1/x^\LCp = 1- \kappa\, \text{Erf}(x^\LCm) \;,
	\ee
	in which the dimensionless variable $\kappa = 2\sqrt{\pi/k} (p_\LCm/(m_0 x^\LCp_0))^2$ in terms of the initial $p_\LCm$. Orbits are plotted for $\kappa\in\{0.3\ldots 0.9\}$. For larger $x^\LCp$ the particles approach the speed of light, as the dynamical mass drops to zero, which makes the Hamiltonian equivalent to that of a massless particle.}
\end{minipage}
\end{figure}

The solution of the equations of motion proceeds as follows. Define $u = x^\LCm - x^\LCperp x^\LCperp /x^\LCp$. Using the conservation of $Q_\LCperp$ and $Q_3$ we may write $p_\LCm$ in terms of $u$,
\be\label{till-pm}
	p_\LCm = - \frac{Q_\LCperp^2 + f(u) }{4Q_3}  \;.
\ee
The Hamiltonian equation of motion for $u$ is, using this,
\be\label{u-DE}
	\frac{\ud u}{\ud x^\LCp} = - \frac{Q_3}{p_\LCm x^{\LCp\, 2}} = \frac{4Q_3^2}{Q_\LCperp^2 + f(u)} \frac{1}{x^{\LCp\, 2}} \implies \int\limits_{u_0}^u \!\ud s\, \frac{Q_\LCperp^2 + f(s)}{4Q_3^2} = \frac{1}{x^\LCp_0} - \frac{1}{x^\LCp} \;, 
\ee
This is an implicit expression for $u\equiv u(x^\LCp)$, with initial conditions $u=u_0$ when $x^\LCp = x^\LCp_0$. From here one identifies $p_\LCm \equiv p_\LCm(x^\LCp)$ via (\ref{till-pm}). The next step is to identify $x^\LCperp$, the Hamiltonian equations of motion for which are
\be
	\frac{\ud x^\LCperp}{\ud x^\LCp} = -\frac{p_\LCperp}{2p_\LCm} =  \frac{x^\LCperp}{x^\LCp} - \frac{Q_\LCperp}{2x^\LCp p_\LCm} \;,
\ee
which can be integrated. With this one finally has an expression for $x^\LCm$ since $x^\LCm = u + x^\LCperp x^\LCperp/x^\LCp$. Example orbits are plotted in Fig.~\ref{FIG:KONF} for the choice of dynamical mass
\be\label{m-def-konf}
	m^2 = \begin{cases}
	m_0^2 & x^\LCp < L \\[5pt]
	\displaystyle  \frac{m_0^2 L^2}{x^{\LCp2}} e^{- k^2 (x^\LCm - x^\LCperp x^\LCperp/x^\LCp)^2} & x^\LCp > L \;,
	\end{cases}
\ee
in which $k$ is a parameter (with units inverse length) and we have turned on the background field at time $x^\LCp=L$.

%%%%%%%%%%%%%%%%%%%%%%%%%%%%%%
\section{Toward the quantum problem}\label{SECT:KG}
%%%%%%%%%%%%%%%%%%%%%%%%%%%%%%
%
%\RefB{////// Work in progress ////// Shorten symmetry discussion. Focus on going from classical to quantum. Point out that the symmetry method is essentially constructive; by mapping the solution to itself we see classical integrability. //////}

\RefB{The quantum mechanical analogues of classical, nonrelativistic, superintegrable systems are obtained by replacing Poisson brackets with commutators~\cite{Marchesiello:2015}. However, relativistic quantum mechanics is problematic~\cite{Klein:1929b}, and the proper framework is relativistic quantum field theory. The `first quantised', quantum mechanical, approach still has a role to play, though; for scalar fields, for example, solutions to the Klein-Gordon (KG) equation give asymptotic particle wavefunctions which are used as the basis of scattering amplitudes. These solutions typically contain, though, physics which cannot be captured by single particle dynamics, the primary example being pair production. It is hence not obvious how the superintegrability of a relativistic particle system translates to its field theory analogue. We will begin to investigate this question here, following the ideas in~\cite{Heinzl:2017zsr} for electromagnetic fields.}

\subsection{\RefB{Scalar fields and symmetry conditions}}
The natural field theory generalisation of our classical system is a quantum field $\varphi(x)$ coupled to an external scalar field $A(x)$ (scalar Yukawa theory, see e.g.~\cite{Heinzl:2018xnv} for recent results and references), with action
\be
	S = \int\!\ud^4 x \  \big(\partial \varphi^\dagger. \partial \varphi - m_0^2 \varphi^\dagger\varphi -A\varphi^\dagger \varphi \big)\;.
\ee
The Euler-Lagrange equations for $\varphi$ yield the KG equation we wish to solve, namely
\be\label{KG}
	\big(\partial^2+m_0^2 +e A(x)\big) \varphi(x) = \big(\partial^2+m^2(x)\big) \varphi(x) = 0\;,
\ee
in which, from here on, the dynamical mass is defined by $m^2(x) \equiv m_0^2+A(x)$.

\RefB{Recall that the linear differential operator $L = \mathcal{L}_\xi + R(x)$, for some transformation $\xi$ and function $R(x)$,  is a symmetry of the KG equation if~\cite{Miller:1977}}
\be\label{general0}
	\big[\partial^2 +m^2 ,  L \big]  = R(x) \big(\partial^2 +m^2\big) \;.
\ee
\RefB{(In other words, $L$ maps solutions of the KG equation to other solutions.)} For $\xi$ in the conformal group considered above it may be checked that \RefB{the following identity holds}:
\be\label{general}
	\bigg[\partial^2+m^2 , \mathcal{L}_\xi +\frac{1}{4}\partial.\xi\, \bigg] = \frac{1}{2}\partial.\xi \big(\partial^2+m^2\big) -  \bigg(\mathcal{L}_\xi m^2 + \frac{1}{2}\partial.\xi\, m^2\bigg) \;.
\ee
\RefB{Interestingly, the final term on the right hand side is precisely as in (\ref{L-xi-general}). When this term vanishes, there is a conserved quantity in the classical particle motion and, here, we see that $L= \mathcal{L}_\xi + \partial.\xi/4$  then becomes a symmetry operator of the KG equation (with $R=\partial.\xi/2)$.} Observe that this may be used \textit{constructively} to solve the KG equation, as follows. We look for solutions to the KG equation which are mapped to \textit{themselves} by the symmetry operator, i.e.~we impose the eigenvector equation
\be\label{KG-villkor}
	\mathcal{L}_\xi \varphi +\frac{1}{4}\partial.\xi\,\varphi = -i Q \varphi \;,
\ee
with $Q$, some constant, the eigenvalue. (A factor of $-i$ is included for convenience.) By solving this eigenvector equation we can \textit{partially} identify the form of a KG solution. 

\RefB{Consider now a superintegrable particle system with $m^2(x)$ obeying (\ref{L-xi-general}), and its associated conformal symmetry generators $\xi$, as in the examples of Sect.~\ref{SECT:EG} above. In order to solve the KG equation with the same $m^2(x)$ we can begin by trying to impose multiple eigenvalue conditions (\ref{KG-villkor}), but only if the corresponding transformations commute. This means that we cannot in general impose all the symmetries of the superintegrable particle system (as, classically, not all of the symmetry generators need be in involution.)} \RefB{Recall, though, the conjecture~\cite{Tempesta:2001} that all maximally superintegrable classical systems are also exactly solvable quantum mechanically. That conjecture was made in the context of non-relativistic systems, so it becomes an interesting question as to what extent it extends to our relativistic case. We have previously investigated several maximally superintegrable systems describing a relativistic particle in an electromagnetic background field, and found in all cases, by imposing the equivalents of (\ref{KG-villkor}) above, that the corresponding Klein-Gordon and Dirac equations can also be solved exactly~\cite{Heinzl:2017zsr}. We therefore proceed to discuss some examples (of Sect.~\ref{SECT:EG} and beyond) in the context of the KG equation. To solve the latter we impose the eigenvector equations (\ref{KG-villkor}) and find in all cases that this reduces the KG equation, a PDE, to a solvable ODE.}

\subsection{Plane waves}
The case of plane waves, $m^2 \equiv m^2(x^\LCp)$, is trivial; imposing (\ref{KG-villkor}) for three translations, $\partial_\LCperp \varphi = - i Q_\LCperp \varphi$, $\partial_\LCm \varphi = - i Q_\LCm \varphi$ for eigenvalues $Q_\LCperp$ and $Q_\LCm$, implies that the solution of the KG equation takes the form
\be
	\varphi = \exp\big(-i Q_\LCperp x^\LCperp -i Q_\LCm x^\LCm\big) \chi(x^\LCp) \;.
\ee
Inserting this into the KG equation leaves a first-order separable ODE for the function~$\chi$,
\be
	\big(\partial^2 + m^2(x^\LCp)\big)\varphi = 0 \implies 4i Q_\LCm \partial_\LCp \chi = \big(p_\LCperp^2 + m^2(x^\LCp) \big)\chi \;,
\ee
and so
\be\label{VolkovSK}
	\varphi = \exp\bigg(-i Q_\LCperp x^\LCperp -i Q_\LCm x^\LCm - i \int\limits^{x^\LCp}\!\ud s\, \frac{p_\LCperp^2 + m^2(s)}{4Q_\LCm}\bigg) \;.
\ee
This is the known general solution to the KG equation in a scalar plane wave, see e.g.~\cite{Heinzl:2018xnv}.

\RefA{The plane wave case gives some insight into the connection with the classical particle results above. Let $S$ be the classical Hamilton-Jacobi function defined by
\be
	\partial_\mu S  = p_\mu \;,
\ee
with $p_\mu$ the classical particle (canonical) four-momentum. Then it is easily verified that the solution (\ref{VolkovSK}) to the Klein-Gordon equation is just the exponential of the classical action, $\varphi=\exp(-i S)$, with $Q_\LCperp = p_\LCperp$ and $Q_\LCm = p_\LCm$ (all three conserved). It follows that for these conserved quantities, and their corresponding generators $\xi$ with $\xi.p=Q$, constant, we can write
\be\begin{split}
	\mathcal{L}_\xi \varphi &\equiv \xi.\partial_\mu \varphi = -i\big(\xi.\partial S \big) \varphi = -i\big(\xi.p  \big) \varphi  = -i Q \varphi \;,
\end{split}
\ee
and thus we see the eigenvector condition (\ref{KG-villkor}) reappearing, through the dependence of the KG solution on the classical Hamilton-Jacobi action, with the conserved particle quantities becoming the eigenvalues.  It would be very interesting to see how this relation holds for the non-Poincar\'e generators in the conformal group, and what one can say in general about the structure of the KG equation.}

\subsection{Special conformal transformations}
We consider again the special-conformal symmetric mass (\ref{KONF-MASSA}), and solve the KG equation. We impose three eigenvalue conditions corresponding to the three classically conserved quantities in involution.  We begin with the conformal transformation $\xi$ with, again, $c^\LCm = 1$ and all other components vanishing, and impose (\ref{KG-villkor}) with eigenvalue $Q_3$,
\be\begin{split}
	\mathcal{L}_\xi \varphi + \frac{1}{4}\partial.\xi \varphi &= -i Q_3 \varphi \\
	\implies \varphi &\equiv  \frac{1}{x^\LCp}\exp\bigg(-i\frac{Q_3}{x^\LCp}\bigg) g\bigg(\displaystyle\frac{x}{x^\LCp},\frac{y}{x^\LCp},x^\LCm - \frac{x^2+y^2}{x^\LCp}\bigg) \;,
\end{split}
\ee
for $g$ an arbitrary function. Imposing (\ref{KG-villkor}) for the two null rotations with eigenvalues $Q_\LCperp$ then fixes the dependence on the first two arguments of~$g$,
\be\label{ppp}\begin{split}
	\big(2 x ^\LCperp \partial_\LCm + x^\LCp \partial_\LCperp \big)\varphi &=  -iQ_\LCperp \varphi \\
	\implies \varphi &\equiv  \frac{1}{x^\LCp}
	\exp\bigg(-i\frac{Q_3+ Q_\LCperp x^\LCperp}{x^\LCp} \bigg)
	g\bigg(\displaystyle x^\LCm - \frac{x^2+y^2}{x^\LCp}\bigg) \;.
\end{split}
\ee
With this we impose the KG equation, with $m^2$ as in (\ref{KONF-MASSA}). The fact that we have already identified much of the structure of $\varphi$, as in (\ref{ppp}) means that the KG equation reduces to a first order ODE in the variable $u \equiv x^\LCm - (x^2+y^2)/x^\LCp$:
\be\begin{split}
	\big(\partial^2+m^2\big)\varphi &= 0 \\
	&\longrightarrow 4 i Q_3\, g'(u) + \big(Q_\LCperp Q_\LCperp + f(u)\big)g(u)  =0 \;.
\end{split}
\ee
It follows that the KG equation is solved \RefB{in separable form} by
\be \label{GWP}
	\varphi(x^\mu) =  \frac{1}{x^\LCp}
	\exp\bigg(-i\frac{Q_3 + Q_\LCperp x^\LCperp}{x^\LCp} +i \int\limits^u \!\ud s\, \frac{Q_\LCperp Q_\LCperp +f(s)}{4Q_3}\bigg) \;,
\ee
as may be verified directly. \RefB{It is known that the KG equation is separable in 261 orthogonal coordinate systems \cite{Kalnins:1978}. Comparing with the literature we note that the solution (\ref{GWP}) is akin to the Gaussian wave packet solution of the two-dimensional Schr\"odinger equation listed in implicit form as the first entry of Table 12 in \cite{Miller:1977}. The closely related separability properties appear natural as the Schr\"odinger equation is obtained from the KG equation via diagonalisation of the derivative $\partial_\LCm$. As a result, both PDEs share a conformal symmetry after the identification of time $t$ in the Schr\"odinger equation with $x^\LCp$ in the KG equation.}

\subsection{Dilations}
%%%%%%
%
As a final example we consider the dynamical mass $m^2(x)\equiv c^2/x.x$ for some constant $c^2$. This mass obeys the symmetry condition (\ref{L-xi-general}) for the two null rotations and dilations (see the third term in (\ref{killing-sol})). These three transformations commute. (We remark that that this mass looks like a generalisation of the non-relativistic $1/r^2$ potential, which is known to exhibit scale symmetry~\cite{Jackiw:1995be}.) The classical system is maximally superintegrable, but we do not present details as the classical orbits have extremely complicated and unrevealing expressions. The field theory case is actually simpler, so we present this instead.

Imposing the eigenvalue condition (\ref{KG-villkor}) for the two null rotations partially identifies the form of $\varphi$, the solution to the KG equation, as 
\be
	\varphi(x) = \exp \bigg(\frac{-iQ_\LCperp x^\LCperp}{x^\LCp}\bigg) g(x^\LCp, x\cdot x) \;,
\ee
where the $Q_\LCperp$ are the two corresponding eigenvalues. Imposing (\ref{KG-villkor}) for dilations with eigenvalue $Q_3$  gives, defining for convenience a new variable $v=\sqrt{x\cdot x}/x^\LCp$,
\be
	\mathcal{L}_\xi \varphi +\frac{1}{4}\partial.\xi \varphi = -i Q_3 \varphi \implies \varphi  = (x^\LCp)^{-(1+i Q_3)}  v^{-i Q_3} \exp \bigg(\frac{-iQ_\LCperp x^\LCperp}{x^\LCp}\bigg) y(v) \;.
\ee
With this, the KG equation reduces to a second-order ODE for the unknown function $y(v)$ in terms of the variable $v$:
\be
	v^2 y''(v)+v y'(v) -\bigg(Q_\LCperp^2 v^2 +c^2- Q_3^2)\bigg)y(v)  = 0 \;.
\ee
This is the defining equation of a Bessel function, with solutions
\be
	y(v) = c_1 J_\alpha(-i|Q_\LCperp|v) + c_2 Y_\alpha (-i|Q_\LCperp|v) \;, \qquad \alpha \equiv \sqrt{c^2 - Q_3^2} \;,
\ee
where the $c_j$ are arbitrary constants.
%%%%%%%%%%%%%%%%%%%%%%%%%%%%%%%
\section{Discussion and Conclusions}\label{SECT:CONCS}
%%%%%%%%%%%%%%%%%%%%%%%%%%%%%%%
%
We have considered the relativistic mechanics of a particle interacting with a background scalar field, or equivalently with a spacetime-dependent mass. Adopting this simplification, relative to the more commonly considered case of a background electromagnetic field, has some advantages. Primarily the lack of vector structure means that the symmetries of the system, giving conserved quantities of the particle motion, are extended from the 10-parameter Poincar\'e group to the 15-parameter conformal group. We have used this to construct several examples of maximally and minimally superintegrable \textit{relativistic} systems.

We have also looked at related results in field theory. The solutions to the Klein-Gordon equation for a scalar field with a dynamical mass provide the asymptotic wavefunctions needed for scattering calculations in (scalar Yukawa) quantum field theory, hence this is a natural `quantum' extension of our relativistic classical particle results. We have found that for the dynamical masses which give superintegrable particle dynamics the Klein-Gordon equation can be solved exactly. Here we have only focussed on giving examples, but it would be very interesting to investigate more systematically how the superintegrability of particle mechanics translates to quantum field theory, especially in light of the conjecture in~\cite{Tempesta:2001}.

\section*{Appendix 1}
The 10-dimensional Poincar\'e group is conveniently parameterised, using the canonical variables of the front form, by
\begin{center}
\begin{tabular}{l|c|c}
Transformation & Notation & $\xi.p$ \\
\hline
4 translations & $p_\mu$ & $p_\LCp$, $p_\LCm$, $p_\LCperp$  \\
1 rotation (about $z$) & $L_z$ & $x p_2 - y p_1$ \\
1 boost  (along $z$) & $K_z$ &  $x^\LCp p_\LCm - x^\LCm p_\LCp$  \\
2 null rotations & $T_\LCperp$ & $2x^\LCperp p_\LCm + x^\LCp p_\LCperp$ \\
2 null rotations & $U_\LCperp$ & $2x^\LCperp p_\LCp + x^\LCm p_\LCperp$
\end{tabular}
\end{center}
% 

%%%%%%%%%%%
\section*{References}
%%%%%%%%%%%

\end{document}